%Paper: astro-ph/9312065
%From: karsten@ulysses.llnl.gov (Karsten Jedamzik)
%Date: Wed, 29 Dec 93 16:27:42 PST

\lineskip=3pt minus 2pt
\lineskiplimit=3pt
\magnification=1200
\centerline{\bf INHOMOGENEOUS PRIMORDIAL NUCLEOSYNTHESIS:}
\vskip 0.05in
\centerline{\bf COUPLED NUCLEAR REACTIONS AND}
\vskip 0.05in
\centerline{\bf HYDRODYNAMIC DISSIPATION PROCESSES}
\vskip 0.5in
\centerline{{\sl Karsten Jedamzik} and {\sl George M. Fuller}}
\vskip 0.15in
\centerline{Department of Physics}
\centerline{University of California, San Diego}
\centerline{La Jolla, CA 92093-0319}
\vskip 0.15in
\centerline{and}
\vskip 0.15in
\centerline{\sl Grant J. Mathews}
\vskip 0.15in
\centerline{Physics Department}
\centerline{University of California}
\centerline{Lawrence Livermore National Laboratory}
\centerline{Livermore, CA 94550}
\vskip 0.3in
\centerline{\bf ABSTRACT}
\vskip 0.08in
\baselineskip=14pt plus 2pt

We present a detailed numerical study of inhomogeneous Big Bang
nucleosynthesis where, for the first time, nuclear reactions
are coupled to all significant fluctuation dissipation processes.
These processes include neutrino heat transport, baryon diffusion,
photon diffusive heat transport, and hydrodynamic expansion
with photon-electron Thomson drag. Light element abundance yields
are presented for broad ranges of initial amplitudes and length
scales for spherically condensed fluctuations. The $^2$H,
$^3$He, $^4$He, and $^7$Li nucleosynthesis yields are found to be
inconsistent with observationally inferred primordial abundances
for all but very narrow ranges of fluctuation characteristics.
Rapid hydrodynamic expansion of fluctuations late in the
nucleosynthesis epoch results in significant destruction of
$^7$Li ($^7$Be) only if the baryonic contribution to the closure
density ($\Omega_b$) is less than or comparable to the upper
limit on this quantity from homogeneous Big Bang nucleosynthesis.
This implies that $^7$Li overproduction will preclude an increase
on the upper limit for $\Omega_b$ for any inhomogeneous
nucleosynthesis scenario employing spherically condensed
fluctuations.
\vskip 0.43in
\centerline{\sl Subject headings: cosmology: early universe -
abundances, nuclear reactions, nucleosynthesis -}
\centerline{\sl hydrodynamics}
\vfil\eject

\baselineskip=24pt plus 2pt
\centerline{\bf 1. Introduction}

In this work we study the primordial nucleosynthesis process in the
presence of non-linear sub-horizon scale baryon-to-photon number
fluctuations. Our study is unique in that we couple all relevant
hydrodynamic dissipation processes to nuclear reactions. It is our
goal to further the understanding of Big Bang nucleosynthesis to the
point where it can be used to probe the fluctuation-generation
physics of the very early universe.

Homogeneous standard Big Bang nucleosynthesis (hereafter, HBBN;
Wagoner, Fowler, \& Hoyle 1967; Wagoner 1973) has become one of the
most important and successful tools for constraining the physics of the early
universe (cf. Schramm \& Wagoner 1977; Yang et al. 1984;
Walker et al. 1991; Smith, Kawano, \& Malaney
1993; Malaney \& Mathews 1993; see also the discussion
in Kolb \& Turner 1990).
Comparision of light element nucleosynthesis yields with
observationally inferred primordial abundances allow limits
to be placed on the baryon-to-photon ratio ($\eta$) and the
expansion rate of the universe and, hence, the energy density
during the nucleosynthesis epoch. One of the key assumptions of
HBBN is that $\eta$ is homogeneously distributed.

By contrast, in inhomogeneous standard Big Bang nucleosynthesis
(IBBN) the assumption of a homogeneous spatial distribution of
$\eta$ is relaxed. In studies of IBBN the object is to calculate primordial
nucleosynthesis abundance yields associated with a given spectrum of
baryon-to-photon number fluctuations extant at the nucleosynthesis
epoch. If this can be done with confidence then the usefulness
of Big Bang nucleosynthesis for constraining the physics of the
early universe can be extended. In particular any process
which generates fluctuations prior to the nucleosynthesis epoch
may be subject to constraint. Possible fluctuation generation
processes associated with the QCD-epoch, the electroweak transition,
and topological defects among others are discussed in the review
by Malaney \& Mathews (1993). Nonlinear fluctuations on the small spatial
scales which could affect primordial nucleosynthesis may even be
produced in an inflationary epoch (cf. Dolgov \& Silk 1992).

The hydrodynamic evolution of fluctuation characteristics from
production at very early times ($T$ $^{<}_{\sim}$ 100 GeV) through
the nucleosynthesis epoch ($T\approx 1$ keV) is discussed in
Heckler \& Hogan (1993) and Jedamzik \& Fuller (1993) - hereafter
referred to as JF93.
In the present study we add to that work a calculation of the light
element nucleosynthesis yields associated with the fluctuation
evolution.

The calculation of primordial nucleosynthesis yields in inhomogeneous
conditions has received extensive previous treatment
(e.g. Epstein \& Petrosian 1975; Applegate, Hogan, \& Scherrer 1987;
Alcock, Fuller, \& Mathews 1987; Malaney \& Fowler 1988;
Fuller, Mathews, \& Alcock 1988; Terasawa \& Sato 1989abc; 1990;
Kurki-Suonio \& Matzner 1989; 1990; Kurki-Suonio et al. 1990;
Mathews et al. 1990). These and other IBBN calculations are reviewed
in Malaney \& Mathews (1993). A vexing problem in computing IBBN abundance
yields
is that dissipative processes which modify fluctuation
characteristics can proceed on time scales which are comparable to
or shorter than nuclear reaction time scales.
The fluctuation damping processes which can be important in the epoch
between $T\approx 1$ MeV and $T\approx 1$ keV are baryon diffusion,
photon diffusive heat transport, and hydrodynamic expansion.
In fact, even the most sophisticated of the calculations listed above
neglect all fluctuation dissipation processes
except for baryon diffusion.

Neglect of photon diffusion and hydrodynamic expansion in previous
IBBN work has left considerable uncertainty in the
nuclear abundance yields
to be expected from given fluctuation characteristics.
Alcock et al. (1990) showed that the
rapid hydrodynamic expansion of fluctuations that sets in after
$e^{\pm}$-pair annihilation conceivably
could result in changes in the $^2$H and $^4$He
yields and in particular destruction of
$^7$Li by orders of magnitude.
This uncertainty in $^7$Li production has diminished the usefulness of IBBN
as a constraint on primordial inhomogeneity.

In this paper we present the results of IBBN calculations which, for the first
time, couple nuclear reactions with photon diffusion, hydrodynamic expansion,
and neutron, proton and ion diffusion. We also include neutrino heat
conduction which modifies fluctuations prior to the nucleosynthesis epoch.
Nucleosynthesis yields are computed for broad ranges of initial
fluctuation characteristics.
These calculations allow us to confidently predict the $^2$H, $^3$He, $^4$He,
and $^7$Li abundances to be expected for particular fluctuation
configurations.

In Section (2) we describe how fluctuations are modeled and numerically
evolved through the nucleosynthesis epoch.
In this section we discuss how fluctuation damping proccesses
are coupled with nuclear reactions in a multi-spatial-zone
calculation.
Section (3) presents the results of our
numerical computations of nucleosynthesis yields. Conclusions are given in
Section (4).

\vskip 0.18in
\centerline{\bf 2. Numerical Fluctuation Evolution and}
\centerline{\bf Nucleosynthesis Calculations}
\vskip 0.18in

We numerically evolve spherically condensed fluctuations from
$T=100$ MeV through the end of the nucleosynthesis era at about
$T\approx 1$ keV.
Initial fluctuation characteristics are specified at the epoch where
$T=100$ MeV. We consider fluctuations in baryon-to-photon number. These
are equivalent to fluctuations in the entropy-per-baryon as discussed in
JF93. We will adopt the notation of that paper here. The amplitude
of a fluctuation in $\eta$ at a position $x$ is defined in terms of the
average net baryon number density in the horizon, $\bar n_b$, and the
appropriate baryonic density $n_i(x)$ as,
$$\Delta_i(x)={n_i(x)\over{\bar n}_b}\ .\eqno(1)$$
This expression represents the amplitude for fluctuations in protons or
neutrons
when the subscript $i$ is replaced by $p$ or $n$, respectively.
Fluctuation amplitudes in light element number density are denoted in a similar
fashion. For example, the spatial distribution of the $^4$He overdensity
is described by $\Delta_{^4He}(x)$. If $\Delta (x)$ has no subscript
it is understood to represent the fluctuation amplitude in total net baryon
number density at position $x$.

We assume that the horizon volume at $T=100$ MeV is filled with a
regular lattice of identical spherical cells, each of which has a
Gaussian shape fluctuation at its center. Although this is certainly
an idealization, the anti-correlated domains expected from, for
example, a homogeneous nucleation of phase, should approximate a regular
lattice. Meyer et al. (1991) have studied the changes in IBBN
abundance yields to be expected for a distribution of fluctuation
centers arising from a realistic nucleation process. In that work
the more realistic distributions give slight increases in the
$^4$He abundance yields. Our calculated abundance yields should be
similarly modified where appropriate.

We refer proper lengths to a
comoving scale at $T=100$ MeV. Consider a length scale
comoving with the Hubble expansion which is measured to have proper
length $L$ at an epoch where the temperature is $T$ and the
scale factor is $R(T)$. The \lq\lq comoving\rq\rq\ length
associated with this length scale $L$ is defined to be $L_{100}$,
$$L_{100}=L\Biggl({R_{100}\over R(T)}\Biggr)\ ,\eqno(2)$$
where $R_{100}$ is the scale factor for the universe at the epoch
where $T=100$ MeV. In what follows we set $R_{100}=1$.

The initial distribution of baryons in the horizon is specified
by giving three fluctuation characteristics:
the amplitude; the Gaussian width; and the
radius of the spherical fluctuation cell. Figure 1 shows a
schematic representation of our fluctuation distribution. In this
figure we plot $log_{10}\Delta(x)$ against position. The Gaussian
width of the fluctuations is $2a_{100}$. The radius of a spherical
fluctuation cell is $l_{100}^s$, so that the mean separation between
fluctuation centers is approximately $2l_{100}^s$.
In our numerical calculations
we characterize the initial amplitude of the
distribution $\Delta (x)$ in a
fluctuation cell by the ratio
$$\Lambda\equiv {\Delta (center)\over\Delta (edge)}\ ,\eqno(3)$$
where $\Delta (center)$ represents the value of $\Delta$ at the
peak of the Gaussian distribution and $\Delta (edge)$ represents
the minimum value of $\Delta$ on the edge of a spherical cell.
Thus
$\Lambda$
is a measure of the density contrast between high and low density
regions of the spherical cell volume.
With these definitions the initial baryon number distribution,
for $\Lambda >>1$,
$$\Delta(x_{100})\approx A\Bigl(\Lambda
exp\bigl(-(x_{100}/a_{100})^2\bigr)+1\Bigr)\ ,\eqno(4a)$$
where $x_{100}$ is the comoving radial coordinate, and $A$ is a
normalization constant determined by,
$$4\pi\int_{0}^{l_{100}^s}x_{100}^2\Delta (x_{100})dx_{100}=1\
.\eqno(4b)$$
In equation (4a) we have assumed that $\Lambda>>1$, but in our
numerical calculations we consider the general case.

Our computations follow the evolution of only one spherical
fluctuation cell.
We employ reflective
boundary conditions at the edge of the spherical cell for all
diffusive and hydrodynamic flows. Reflective boundary conditions are
appropriate in the limit where the fluctuation cells form a regular lattice
(cf. Figure 1).

We divide a spherical fluctuation cell into 24 spatial zones.
We solve for the hydrodynamic evolution of the fluctuation cell in
the manner described in JF93 but with special modifications for
treating coupled nuclear reactions and photon dissipative
processes.
Baryons, ions, and heat are
transported between zones by radiative conduction of neutrinos and
photons, diffusion, and hydrodynamic flows.
We also follow the annihilation of $e^{\pm}$-pairs in
detail. However, the effects of muon and pion annihilation are neglected
in these computations. This approximation is justified in the
discussion in JF93.

Nuclear abundances and the energy density in each zone can be
modified by nuclear reactions. We include the effects of nuclear
reactions in our evolution calculations by solving a reaction
network in each spatial zone. The reaction network in each zone is
coupled to the nuclear evolution in the other zones through baryon
and ion diffusion and hydrodynamic processes. Nuclear reactions in
each zone are followed in a manner similar to that of the
Wagoner (1967; 1973) code as updated by Kawano (1992). The reaction
network in our calculation includes nuclei up to and including mass
$A=12$. The reaction rates employed are somewhat modified from the
Smith, Kawano, \& Malaney (1993) study and are discussed in Section
3.

As in Wagoner et al. (1967)
we employ a second order Runge-Kutta driver
to follow the time evolution of the nuclear
abundances
${\rm Y}_i^k$ ($i$ - nuclide index,
$k$ - zone index), the scale factor $R$, and the cosmic average
temperature $T$.
The ${\rm Y}_i^k$ are number fractions of nuclear species $i$ relative to
hydrogen (protons) in zone $k$.
In the Runge-Kutta algorithm, changes in nuclear abundances and
thermodynamic quantities are computed twice and averaged for each
time step.

We employ a fully implicit differencing scheme for the simultaneous
coupling of nuclear reactions and neutron diffusion. In this scheme
we invert a matrix of nuclear reaction rates and zone-to-zone
neutron diffusion rates for each time step. Neutron diffusion and
nuclear reactions must be coupled completely implicitly in order to
follow the nuclear evolution of high amplitude fluctuations. This is
because nuclear reaction and neutron diffusion time scales
can become comparable for these fluctuations.
Neutron diffusion rates for smaller amplitude fluctuations are
usually not as fast as typical nuclear reaction rates.
A coupling scheme for high
amplitude fluctuations which employs only a partially
implicit algorithm requires decreasing time steps. This
may result in a
time-step-crash. In our calculations we find that an explicit
coupling of proton and ion diffusion with nuclear reactions is
sufficient to follow the evolution of nuclear abundances accurately.

Photon diffusive heat transport, hydrodynamic expansion with
radiation drag, and nuclear reactions are adequately modeled with an
explicit coupling in our numerical computations. The physics of
fluctuation modification by photon diffusion and hydrodynamic
expansion is studied in JF93. In that work it is shown
that a key quantity for determining the efficacy of these processes
is the photon mean free path $l_{\gamma}(x,t)$. Photons can be
either optically thick or optically thin across a spatial zone,
according to whether $l_{\gamma}(x,t)$ (evaluated at the zone center)
is smaller than or larger than the zone size.

Our numerical calculations employ a diffusive photon heat transport
scheme when photons are optically thick. Zones coupled in this
manner are well approximated as being in pressure equilibrium with
each other. The effects of photon diffusive heat transport
during a time step are modeled by expansion or contraction of
individual zones to obtain a new pressure equilibrium. In this
manner zones evolve through a succession of pressure equilibrium
states. This scheme will be accurate as long as the sound crossing
time is short compared to the photon diffusive heat transport time
between zones. We employ a Lagrangian
grid to facilitate the modeling of expansion or contraction of
individual zones.

In the limit where photons are optically thin the approximation of
pressure equilibrium between zones breaks down. In this case
pressure gradients result in local acceleration of fluid elements.
Fluid velocities rapidly obtain terminal velocities due to
photon-electron Thomson drag (Alcock et al. 1990; JF93).
In our IBBN calculations we neglect acceleration times and treat
terminal velocities as being attained instantaneously. We determine
terminal fluid velocities by balancing local
pressure-gradient-induced hydrodynamic stresses against radiation
drag forces. In our calculations hydrodynamic expansion is modeled
with an Eulerian grid which allows for fluid motions across zone
boundaries.

In the calculations of fluctuation evolution it is common to
encounter a situation in which the high density central zones are
optically thick to photons while the low density outer zones are
optically thin. An exact treatment of the momentum transfer between
baryons and photons in this case would require a solution of the
Boltzmann equation. In this study we approximate the transition between
optically thick and thin transport regimes by using a superposition
of the techniques outlined above for the two limiting cases.

We introduce an independent set of \lq\lq photon zones\rq\rq .
These zones are required to be larger than the photon mean free path
at any point and time. The photon zones and baryon zones are taken
to be initially coincident at $T=100$ MeV. For lower temperatures
the photon mean free path can grow rapidly, especially during the
$e^{\pm}$-pair annihilation epoch. This necessitates employing a
re-zoning procedure to produce increasingly larger photon zones.
Eventually photon zones and baryon zones are not coincident. For
example, one photon zone could include a cluster of baryon zones,
each having different baryon content and nuclear abundances but the same
photon temperature. In regimes where rapid hydrodynamic expansion
occurs the entropy generation due to radiation drag is rapid enough
to justify an assumption of uniform photon, electron, and baryon
temperature.

Where baryon and photon zones are not coincident the numerical
procedures for computing fluid velocities and pressure equilibrium
conditions must be chosen carefully in order to avoid time step and
spatial instabilities. In our IBBN calculations local fluid
velocities are determined by an effective pressure gradient between
adjacent baryon zones. This effective pressure gradient includes
contributions from gradients in effective photon temperature as well
as from gradients in neutron, proton, ion, and electron pressures. In the case
where photon zones contain many baryon zones we assign effective
photon temperatures to each baryon zone. These effective
temperatures are assigned by interpolating between photon
temperatures in adjacent photon zones. In this procedure we allow
for fluid movements across common baryon and photon zone boundaries.

We employ a three step algorithm for computing the effects of fluid
flow, heat transport, and radiation drag on fluctuation evolution.
First we compute changes in energy density in photon zones arising
from photon heat diffusion. This causes deviations from pressure
equilibrium between photon zones. In the second step we compute the
effects of local effective pressure gradients and associated fluid
movements across baryon zone boundaries. Finally we expand or
contract baryon and photon zones to achieve pressure equilibrium
where appropriate.

In order to obtain optimal convergence in nucleosynthesis yields we
place most of the baryon zones where the baryon number gradients are
largest. Figure 2 displays the average abundance yields in $^4$He
(solid line) and $^7$Li (dashed line) as a function of number of
baryon zones employed in a particular calculation. This calculation
evolved a
fluctuation which was taken to have large initial density contrast
($\Lambda = 1.25\times 10^6$). Convergence for such high amplitude fluctuations
is problematic due to the pronounced effects of hydrodynamic
damping.
In this investigation we use 24 baryon zones, implying
intrinsic computational uncertainties in the nucleosynthesis yields
of $\Delta {\rm Y}_p/{\rm Y}_p\sim 1\%$,
$\Delta {\rm Y}_{{}^2{\rm H}}/{\rm Y}_{{}^2{\rm H}}\sim 10\%$,
and $\Delta {\rm Y}_{{}^7{\rm Li}}/{\rm Y}_{{}^7{\rm Li}}\sim 20\%$.
Here ${\rm Y}_{{}^2{\rm H}}$ and ${\rm Y}_{{}^7{\rm Li}}$ denote number
fractions relative to hydrogen, while ${\rm Y}_p$ represents the
primordial $^4$He
mass fraction.

\vskip 0.18in
\vfil\eject
\centerline{\bf 3. Results of Inhomogenous}
\centerline{\bf Big Bang Nucleosynthesis Calculations}
\vskip 0.18in

In this section we discuss the nucleosynthesis yields obtained with
our coupled nuclear reaction and hydrodynamic damping simulations.
In general, our calculated average nucleosynthesis yields agree well
with those obtained by Kurki-Suonio \& Matzner (1989; 1990);
Kurki-Suonio et al. (1990); Terasawa \& Sato (1989abc); and
Mathews et al. (1990) when the fluctuations considered are
relatively low in initial density contrast ($\Lambda$) and have
large initial volume filling fractions (i.e. small $l_{100}^s/a_{100}$).
This is not surprising, as we would expect photon diffusive effects
and hydrodynamic damping to be minimal for fluctuations with these
characteristics. By contrast, our results for fluctuations with high
$\Lambda$ and large $l_{100}^s/a_{100}$ can differ significantly
from previous studies. We therefore concentrate the discussion here
on the nucleosynthesis yields obtained from fluctuations with
larger values
for $\Lambda$ and $l_{100}^s/a_{100}$. We also consider a wider range in
the fractional contribution of baryons to the closure density
($\Omega_b$) than do previous IBBN studies.

Figures 3, 4, 5, 6, and 7 show the fluctuation amplitude ($\Delta_i$)
as a function of cell radius $r_{100}$ for protons (solid line),
neutrons (dotted line), $^2$H (long-dashed line), $^3$He
(short-dashed-dotted line), $^4$He (short-dashed line),
$^7$Li (long-dashed-dotted line), and $^7$Be (short-dashed line).
Confusion between the lines for $^4$He and $^7$Be can be avoided by
noting that the abundance of $^4$He is always orders of magnitude
larger than that for
$^7$Be. The total $^7$Li yield long after the nucleosynthesis epoch
is the sum of the $^7$Li and $^7$Be abundances shown on these plots.
Each figure is labeled by the initial values of $\Lambda$,
$l_{100}^s$, and $l_{100}^s/a_{100}$, and the overall average
$\Omega_b$ used in the calculation. In the expressions for
the baryonic contribution to the closure density
$h$ is the value of the Hubble parameter in units of
100 km s$^{-1}$Mpc$^{-1}$. The upper limit on $\Omega_b$ from
homogeneous primordial nucleosynthesis is
$\Omega_b\approx 0.013h^{-2}$.

Figure 3 shows nine \lq\lq snapshots\rq\rq\ in the evolution of a
fluctuation with initial characteristics
$\Lambda =3\times 10^7$, $l_{100}^s=0.5$ m, and
$l_{100}^s/a_{100}=25$. We use an overall average baryonic density
of $\Omega_b =0.025h^{-2}$ in this calculation. The snapshots
correspond to epochs with temperatures (in MeV)
T=100, 5.0, 2.5, 1.0, 0.25, 0.10, 0.05, 0.025, 0.01.
Figures 4 through 7 show snapshots in the nucleosynthesis yields and
fluctuation amplitude evolution for different initial fluctuation
parameters and values of $\Omega_b$. In these figures the snapshots
correspond to epochs with temperatures (in MeV) T=100, 1.0, 0.25,
0.10, 0.075, 0.050, 0.025, 0.01, 0.008. Figure 4 shows the evolution
of a fluctuation with initial characteristics $\Lambda =3\times
10^7$, $l_{100}^s=50$ m, and $l_{100}^s/a_{100}=25$. This evolution
sequence is computed with an overall average $\Omega_b=0.25h^{-2}$.
Figure 5 is the same as Figure 4 but with $\Omega_b=0.0125h^{-2}$
and with initial fluctuation characteristics $\Lambda =1.25\times
10^6$, $l_{100}^s=14$ m, and $l_{100}^s/a_{100}=20$. Figure 6 is
similar to Figure 4 but with parameters $\Omega_b=0.25h^{-2}$,
$\Lambda =2\times 10^9$, $l_{100}^s=50$ m, and
$l_{100}^s/a_{100}=100$. Similarly Figure 7 shows the fluctuation
evolution for parameters $\Omega_b=0.0083h^{-2}$, $\Lambda
=1.25\times 10^7$, $l_{100}^s=50$ m, and $l_{100}^s/a_{100}=1000$.
Figures 6 and 7 show the various fluctuation amplitude profiles only
for the core regions of the fluctuation cells.

Average nuclear abundance yields computed for a variety of initial
fluctuation characteristics and overall average values of $\Omega_b$
are given in Figures 8, 9, 10, 11, 12, 13, and 14.

Figures 8, 10, 12, and 14 show average nucleosynthesis yields of
$^4$He, $^7$Be (unstable), $^7$Li, $^2$H, $^3$He, and $^3$H
(unstable) as a function of fluctuation cell radius $l_{100}^s$ for
various values of average $\Omega_b$, $l_{100}^s/a_{100}$, and
$\Lambda$. Figure 8 corresponds to parameters $\Omega_b =0.25h^{-2}$,
$l_{100}^s/a_{100}=25$, and $\Lambda =3\times 10^7$. Figure 10
corresponds to $\Omega_b=0.025h^{-2}$, $l_{100}^s/a_{100}=20$,
and $\Lambda =1.25\times 10^6$; whereas, Figure 12 corresponds to parameters
$\Omega_b=0.0125h^{-2}$, $l_{100}^s/a_{100}=20$, and $\Lambda
=1.25\times 10^6$; and Figure 14 corresponds to parameters
$\Omega_b=0.005h^{-2}$, $l_{100}^s/a_{100}=20$, and $\Lambda
=1.25\times 10^6$. These figures also give the average
$^7$Be abundance
(at $T=20$ keV and 5 keV) and the average neutron-to-proton ratio
(at $T=20$ keV) as a function of initial fluctuation cell radius
$l_{100}^s$.

Figures 9, 11, and 13 show average nucleosynthesis yields for the
same light elements as in Figures 8, 10, 12, and 14 but now presented
as a function of initial fluctuation Gaussian width $a_{100}$.
The calculations presented in Figure 9 use $\Omega_b=0.25h^{-2}$,
$l_{100}^s=50$ m, and employ a large density contrast ($\Lambda >10^6$).
The results shown are insensitive to $\Lambda$ for $\Lambda >10^6$.
Figure 11 corresponds to parameters $\Omega_b=0.025h^{-2}$,
$l_{100}^s=10$ m, and $\Lambda =1.25\times 10^7$; whereas,
Figure 13 corresponds to parameters $\Omega_b=0.0125h^{-2}$,
$l_{100}^s=24$ m, and $\Lambda =1.25\times 10^7$.

Several important general trends and results are evident in our IBBN
calculations. We find that fluctuations whose cell
center separations satisfy $2l_{100}^s$ $^<_{\sim}$ 1 m are
effectively damped to near homogeneity by the onset of nuclear
freeze-out at $T\approx 100$ keV (see Figure 3).
Comparision of observationally inferred light element primordial
abundances with IBBN calculations can only be used to constrain
primordial baryon-to-photon number fluctuations with $2l_{100}^s$
$^>_{\sim}$ 1 m. This conclusion is relatively
insensitive to the value of
$\Omega_b$ and initial values of $\Lambda$ and $a_{100}$.
Primordial nucleosynthesis in a universe with fluctuations whose
separations are smaller than $2l_{100}^s\approx 1$ m is essentially
identical to HBBN at the same average $\Omega_b$.

We do not find substantial late-time destruction of $^7$Be ($^7$Li)
as a result of hydrodynamic expansion for any
fluctuation characteristics
when $\Omega_b$ $^>_{\sim} 0.025h^{-2}$. This is in contrast to
suggestions based on simple mixing calculations reported in
Alcock et al. (1990) and the general schemes for $^7$Be ($^7$Li)
destruction by neutron back diffusion proposed by Malaney and
Fowler (1988) for high average $\Omega_b$ universes. Our results
imply
that any spherically condensed fluctuation characteristics
will produce average ${}^7{\rm Li/H}$ $^>_{\sim} 10^{-9}$ whenever
$\Omega_b > 0.013h^{-2}$. For some extreme fluctuation
characteristics we find ${}^7{\rm Li/H}\approx 10^{-9}$
even for very large
$\Omega_b$, but in these cases $^4$He is overproduced. These
considerations force us to surmise that
{\it any} IBBN scenario
employing spherically condensed fluctuations will yield essentially
the same upper limit on $\Omega_b$ as HBBN.

For low $\Omega_b$ (see Figures 12, 13, and 14 where
$\Omega_b\leq 0.0125h^{-2}$) there can be substantial
hydrodynamic-expansion-induced late-time destruction of $^7$Li
(as $^7$Be). For a narrow range of fluctuation parameters we
find ${}^7{\rm Li/H}\approx 1-3\times 10^{-10}$, with
the $^2$H, $^3$He, and $^4$He yields meeting observational
constraints.
This low $^7$Li nucleosynthesis yield is a result of diffusive
and hydrodynamic dissipation effects {\it during} the
nucleosynthesis era.
It is possible to find fluctuation characteristics (Figure 14) for
which the average $^4$He mass fraction ($Y_p\approx 0.225$)
is below the minimum from HBBN
calculations, but at the cost of a slight overproduction in $^2$H
(${}^2{\rm H/H}\approx 2.5\times 10^{-4}$) and $^7$Li
(${}^7{\rm Li/H}\approx 2.5\times 10^{-10}$).
The recent re-estimate of the ${}^7{\rm Li}({}^2{\rm H},n)2\alpha$
rate by Boyd, Mitchell, \& Meyer (1993) indicates that this process
may play an important role in late-time $^7$Li destruction. We have
not incorporated this new rate into our calculations, but its effect
may be to decrease our $^7$Li yields by about a factor of two.
We discuss the effects of dissipation processes and nuclear reaction
uncertainties on $^7$Li IBBN yields in Section 3.2.

The studies by Heckler \& Hogan (1993) and JF93 show that
fluctuations with a broad range of characteristics, but possessing
initially large amplitudes, will have convergent evolution. These
fluctuations are damped efficiently by neutrino heat conduction at
high temperatures ($T\approx 50-80$ MeV) and converge to a generic
amplitude ($\eta\sim 10^{-4}$) and shape. This is illustated in
Figure 6.

Fluctuations with these characteristics can produce a substantial
mass fraction in elements heavier than $A>7$ ($X_{A>7}\sim 10^{-9}$).
If only a small fraction of the baryons ($^<_{\sim}$ 2\%) reside in
the high density cores of these fluctuations then all light element
abundance constraints can be satisfied.
Figure 7 shows the evolution of a fluctuation with the appropriate
characteristics.
Jedamzik et al. (1993)
have used an extended nuclear reaction network to investigate
the abundance pattern which emerges from the nucleosynthesis process
in such fluctuations. They find that intermediate mass nuclei
on the proton-rich side of the valley of beta stability can be
significantly produced as a possible observable signature of such
inhomogeneities.

Our calculations show that late-time r-process production of heavy
nuclides associated with the neutron-rich low-density regions of
fluctuation cells (Applegate 1988; Applegate et al. 1993)
is very unlikely. We find that neutron back
diffusion effectively eliminates neutron-rich low-density
environments before the onset of an r-process is possible.

Our investigation shows that it is remarkably difficult to reconcile
inhomogeneous Big Bang nucleosynthesis yields with
observationally-inferred primordial light element abundances.
In fact only very narrow ranges of fluctuation characteristics
can give nuclear abundances which satisfy observational constraints.
In the following subsections we discuss some specific features
of IBBN production of $^4$He, $^7$Li, $^2$H, and $^3$He.

\vskip 0.15in
\centerline{\bf 3.1. $^4$He}
\vskip 0.08in

The production of $^4$He in IBBN models can be sensitive to neutron
back diffusion but is influenced very little by late-time
hydrodynamic fluctuation damping. In the proton-rich high-density
core regions of fluctuation cells $^4$He is synthesized in
essentially the same manner as in HBBN. In these regions nearly all
of the neutrons which remain at the epoch of nuclear statistical
equilibrium (NSE) freeze-out are incorporated into alpha particles.
This implies that the neutron number density in these regions serves
as a bottleneck for $^4$He production. Substantial $^4$He production
in the
high-density cores in our models occurs at temperatures between
$T\approx 300$ keV and $T\approx 100$ keV, depending upon the
fluctuation characteristics.

The low-density outer regions of fluctuation cells can be
neutron-rich. The synthesis of $^4$He in these regions generally
proceeds at later times and lower temperatures than in higher
density proton-rich regimes. Significant $^4$He synthesis
in these regions
typically occurs for temperatures $T\sim 50$ keV.

The high and low density regions in fluctuation cells may or may not
be efficiently coupled by neutron diffusion. Neutrons cannot readily
diffuse out of high density regions if fluctuation centers have
separations ($2l_{100}^s$) which are very large compared to the
neutron diffusion length at NSE freeze-out and have smaller values
of $l_{100}^s/a_{100}$. The result is $^4$He overproduction.
This trend is evident in Figure 4. Likewise, if fluctuations have
values of $2l_{100}^s$ which are small compared to the neutron
diffusion length then efficient back diffusion of neutrons from low
density neutron-rich regions into higher density zones takes place.
This back diffusion is driven by the gradient in neutron number
density which develops as neutrons are consumed in the higher
density regions. Most neutrons which diffuse back toward high
density zones are incorporated into alpha particles. The result can
again lead to $^4$He overproduction. This is evident in Figure 5,
where back-diffusion-driven $^4$He production is dominant.

The average nucleosynthesis yield of $^4$He can be a minimum when
the value of $2l_{100}^s$ is roughly the same as the neutron
diffusion length at the epoch of NSE freeze-out. This $^4$He
\lq\lq dip\rq\rq\ is a general feature of IBBN models
(Alcock, Fuller, \& Mathews 1987; Kurki-Suonio et al. 1988;
Mathews et al. 1990).
Figures 10, 12, and 14 give good examples of the $^4$He dip. Note
that the position of the dip depends upon fluctuation characteristics
to some extent. In our studies the dip occurs for $l_{100}^s$
between 20 m and 50 m and becomes deeper and more pronounced
for lower values of $\Omega_b$.

Photon diffusive damping and hydrodynamic expansion of fluctuations
only become significant after the epoch of $e^{\pm}$-annihilation.
The photon opacity depends sensitively on the number density
of electrons and positrons. Pairs make an appreciable contribution
to the total electron number density until the temperature falls
below about $T\approx 30$ keV to $T\approx 20$ keV. Hydrodynamic
effects are important when the photon mean free path becomes
comparable to the size of the high-density core regions of
fluctuations.
In our studies this typically occurs in
approximate coincidence with the end of $e^{\pm}$-pair-domination
of the electron number density. At these temperatures $^4$He
production is inhibited by the Coulomb barriers of key reaction
rates. This conclusion does not apply to late-time $^7$Be ($^7$Li)
destruction or $^2$H production.

\vskip 0.15in
\centerline{\bf 3.2. $^7$Li}
\vskip 0.08in

In IBBN models $^7$Li can be produced directly by
${}^3{\rm H}(\alpha ,\gamma ){}^7{\rm Li}$ in neutron-rich low-density
conditions, or as $^7$Be via ${}^3{\rm He}(\alpha ,\gamma ){}^7{\rm
Be}$ in high-density regions.
The $^7$Be is converted to $^7$Li by
${}^7{\rm Be}(e^-,\nu_e){}^7{\rm Li}$ on a time scale long compared
to nucleosynthesis times.

Most of the total average $^7$Li yield comes from the $^7$Be
production channel whenever fluctuations have large $\Lambda$ and small
$l_{100}^s/a_{100}$. This is evident in Figures 4 and 6. These
figures show that a significant fraction of the $^7$Be can be produced in a
spherical
shell at the edge of the high-density core region in a fluctuation
cell. This spatial distribution of $^7$Be makes the average
nucleosynthesis yields of $^7$Li in IBBN models particularly
sensitive to back diffusion (Malaney \& Fowler 1988).
Back diffusion of neutrons into the $^7$Be shell can induce the
reaction sequence ${}^7{\rm Be(n,p)}{}^7{\rm Li(p,}\alpha )\alpha$.
These reactions can reduce the ultimate average $^7$Li
nucleosynthesis yield in IBBN models. This destruction process can
be very effective due to the extremely large value of the $^7$Be
neutron capture cross section.

Hydrodynamic expansion of the high density core regions of
fluctuations can effectively enhance the back diffusion of neutrons
and the associated destruction of $^7$Be. Alcock et al. (1990)
modeled the fluctuation expansion by instantaneously mixing high and low
density material at temperature $T_m$. This mixing temperature was
taken to coincide with the beginning of rapid photon-drag-limited
expansion. They found, however, that the efficiency of the $^7$Be
destruction process was sensitive to the choice of $T_m$ and, hence,
to particular fluctuation characteristics.

In Figure 15 we compare the yields from our hydrodynamic
calculations with simple mixing approximations. In this figure we
plot the average $^7$Be/H in a fluctuation cell against cosmic
temperature. The initial fluctuation parameters employed in this
calculation are $\Lambda =2\times 10^7$, $l_{100}^s=50$ m,
and $l_{100}^s/a_{100}=20$. In this calculation we assumed a large
value for the baryonic density, corresponding to
$\Omega_b=0.25h^{-2}$. This figure shows a fully hydrodynamic
calculation result (solid line), a case where hydrodynamic expansion
times are decreased by a factor of 20 (dotted line), and three
schematic mixing calculations where all nuclear species and densities are
instantaneously homogenized at the indicated value of $T_m$. The
dashed-dotted line corresponds to a calculation with mixing at
$T=27$ keV in which Reaction Rate Set I is employed (Table 1).
The short-dashed and long-dashed curves correspond to mixing
calculations with $T_m=20$ keV in which we used Reaction Rate Sets
I and II, respectively. Table I lists reaction rates for
${\rm p(n,}\gamma ){}^2{\rm H}$ and ${}^7{\rm Be(n,p)}{}^7{\rm Li}$
taken from Smith et al. (1993), Mathews et al. (1990), and
Fowler (1993). The Smith et al. (1993) rates constitute Reaction
Rate Set
I, while the Mathews et al. (1990) and Fowler (1993) comprise
Reaction Rate Sets II and III, respectively. Our hydrodynamic
calculations employ Reaction Rate Set II.

The mixing calculations shown in Figure 15 reproduce the
results found in Alcock et al. (1990). We note, however, that our
realistic hydrodynamic calculations do not show the large amounts of
$^7$Be destruction obtained in mixing calculations.
Even if we artificially increase the
expansion rate of exploding fluctuations by a factor of 20
this conclusion would not change.
Figure 15 shows that $^7$Be late-time destruction can be sensitive
to the nuclear reaction rates employed in the calculations.
This fact can be important for IBBN calculations at low $\Omega_b$,
where $^7$Be destruction is more significant.

General considerations of neutron numbers and reaction cross
sections can be used to understand the requirements for significant
$^7$Be destruction. Two processes which remove neutrons compete
with the ${}^7{\rm Be(n,p)}{}^7{\rm Li}$ destruction channel:
radiative neutron capture by protons; and free neutron decay.
Neutrons diffusing back into high density regions will be consumed
principally by either ${\rm p(n,}\gamma ){}^2{\rm H}$ or
${}^7{\rm Be(n,p)}{}^7{\rm Li}$. Comparing the reaction rates for
these processes we find,
$$\Biggl({1\over {\rm Y_n}} {d{\rm Y_n}\over dt} \bigg|_{\rm pn}\Biggr)
\Bigg/\Biggl( {1\over {\rm Y_n}} {d{\rm Y_n}\over dt} \bigg|_{\rm nBe}
\Biggr)={{\lambda_{\rm pn}{\rm Y_H}}\over {\lambda_{\rm nBe}{\rm Y}_
{{}^7{\rm Be}}}}\approx 6\times 10^{-6}{{\rm Y_H}\over {\rm
Y}_{{}^7{\rm Be}}}\ ,\eqno(5)$$
where ${\rm Y_n}$, ${\rm Y_H}$, and ${\rm Y}_{{}^7{\rm Be}}$ are the
neutron, proton and $^7$Be number fractions relative to hydrogen,
respectively. With this notation ${\rm Y_H}=1$. In equation (5)
$\lambda_{\rm pn}$ is the reaction rate for ${\rm p(n,}\gamma
){}^2{\rm H}$, while $\lambda_{\rm nBe}$ is the rate for
${}^7{\rm Be(n,p)}{}^7{\rm Li}$. A $^7$Be abundance of
${\rm Y}_{{}^7{\rm Be}}\sim 10^{-8}$
is typical for the epochs where rapid hydrodynamic
expansion is likely.
{}From this and equation (5) we see that only about one in 600 neutrons will be
captured on $^7$Be. We conclude that efficient $^7$Be destruction
would require that roughly 600 neutrons per $^7$Be nucleus be
delivered to the high density core regions of fluctuations by
neutron back diffusion. This neutron number requirement might be
lowered somewhat if the $^7$Be resides in a narrow shell on the edge
of the high density region.
Requiring 600 neutrons per $^7$Be nucleus in order to have efficient
$^7$Be destruction implies that the average neutron abundance has to
exceed $Y_n$ $^>_{\sim}$ $600Y_{^7{\rm Be}}\approx 6\times
10^{-6}\sim 10^{-5}$.
This condition is independent
of the $^7$Be abundance.
Free neutron decay will reduce
${\rm Y_n}$ below $10^{-5}$ unless hydrodynamic expansion and the
associated enhancement in the neutron back diffusion rate occur
early enough ($T$ $^>_{\sim} 20$ keV).

Figures 10 and 12 illustrate how the $^7$Be/H (${\rm Y}_{{}^7{\rm
Be}}$) and neutron-to-proton
ratios depend on $l_{100}^s$ for representative fluctuation
characteristics. It is clear from these figures that ${\rm
Y_n}>10^{-5}$ when $l_{100}^s$ $^>_{\sim}$ 10 m at an epoch where
$T=20$ keV. We note that appreciable $^7$Be destruction occurs for
$l_{100}^s$ between 10 m and 50 m, as expected. For $l_{100}^s>50$ m
we find ${\rm Y_n}>10^{-5}$ but late-time expansion is inefficient
and there is little $^7$Be destruction.

The efficiency of $^7$Be destruction associated with late-time
hydrodynamic disassembly of fluctuations depends upon the average
value of $\Omega_b$ employed in the calculations. In Figure 16 we
show the $^7$Li and $^7$Be average abundances as a function of
cosmic temperature derived from calculations with hydrodynamic
expansion (solid curve) and without (dashed curves). The fluctuation
parameters here are the same as in Figure 10 except that $l_{100}^s$
is fixed at $l_{100}^s=12$ m. These calculations employ a large
value for the average baryon density, $\Omega_b=0.025h^{-2}$.
The overall
$^7$Li yield in this case is dominated by $^7$Be production in
higher density regions. We see that hydrodynamic effects induce a
decrease in $^7$Be/H by a factor of 1.5. Note that
a small amount of $^7$Li is
produced in the course of the $^7$Be destruction. This is because the
${}^7{\rm Li(p,}\alpha )\alpha$ rate is not completely effective in
destroying $^7$Li in the high density zones at low temperatures.

For low average $\Omega_b$ the $^7$Be destruction caused by
hydrodynamic fluctuation damping can be significant. Figure 17 shows
$^7$Li and $^7$Be average abundances as a function of temperature
for calculations with and without hydrodynamic effects.
The notation in this figure is the same as in Figure 16.
An average baryon density of $\Omega_b=0.0125h^{-2}$ was used in
the calculations whose results are shown in Figure 17. This
$\Omega_b$ is a factor of two lower than that employed in the
calculation shown in Figure 16. The fluctuation parameters employed
for Figure 17 are the same as those in Figure 16 except for an
insignificant difference in $l_{100}^s$.
In the low $\Omega_b$ calculation we see that $^7$Be is
destroyed by approximately a factor of 5.
However, in this case most of the final average $^7$Li yield comes
from direct $^7$Li production in low density regions.

We find that lower average $\Omega_b$ leads to an earlier onset
time and an enhanced vigor for hydrodynamic fluctuation dissipation.
These effects result in increased $^7$Be destruction. The onset of
rapid hydrodynamic evolution occurs when the comoving photon mean
free path, $l_{100}^{\gamma}$, exceeds the comoving length scale
associated with the high density core region of a fluctuation cell,
$L_{100}^H$. Our calculations show that,
$$\biggl({l_{100}^{\gamma}\over L_{100}^H}\biggr)\approx 0.4\bigl(
\Omega_{eff}^H\bigr)^{-1}\biggl({T\over 20 {\rm keV}}\biggr)^{-2}
\biggl({L_{100}^H\over {\rm m}}\biggr)^{-1}\ ,\eqno(6)$$
where $\Omega_{eff}^H$ is the effective baryon density in the core
region divided by the closure density. For given fluctuation
characteristics a lower average $\Omega_b$ will result in a smaller
$\Omega_{eff}^H$ and, therefore, an earlier onset time for hydrodynamic
expansion. A smaller value for $L_{100}^H$ produces a similar
effect. Figure 13 illustrates these trends. In interpreting this
figure it helps to note that small $a_{100}$ tends to
be associated with smaller
values for $L_{100}^H$. This figure shows a nearly two order of
magnitude destruction of $^7$Be for small $a_{100}$.

\vskip 0.15in
\centerline{\bf 3.3. Deuterium and $^3$He}
\vskip 0.08in

The nucleosynthesis yield of $^3$He in our IBBN study is
relatively insensitive to fluctuation parameters. This is because
$^3$He is produced in both high and low density regions of a
fluctuation cell. As is evident from Figures 8 through 14 we find a
$^3$He abundance relative to hydrogen of
$2\times 10^{-6} < {\rm Y}_{{}^3{\rm He}} < 3\times 10^{-5}$ in our
survey. Large $^3$He abundances are produced only in low average
$\Omega_b$ universes.

Deuterium is produced only in relatively low density regions of
fluctuation cells. We find that average $^2$H abundance yields are
sensitive to $l_{100}^s$, $l_{100}^s/a_{100}$, and $\Lambda$.
There is a rather firm lower bound on the primordial $^2$H abundance
of ${\rm Y}_{{}^2{\rm H}}$ $^>_{\sim} 1.8\times 10^{-5}$
(Walker et al. 1991). This limit presupposes no significant
non-Big-Bang source of $^2$H (see the discussion in Gnedin \&
Ostriker 1992). Dearborn et al. (1986) derive an upper bound on the
sum of $^2$H and $^3$He of
$({\rm Y}_{{}^2{\rm H}}+{\rm Y}_{{}^3{\rm He}})$ $^<_{\sim}$ $9\times
10^{-5}$. We find that IBBN models with low average $\Omega_b$
frequently produce $^2$H yields which exceed this upper limit.

\vskip 0.18in
\centerline{\bf 4. Conclusions}
\vskip 0.18in

We have studied the primordial nucleosynthesis process for universes
with spherically-condensed high-amplitude baryon-to-photon
fluctuations. Unlike previous efforts to investigate this problem, our
numerical calculations couple nuclear reactions with all important
diffusive and hydrodynamic fluctuation dissipation mechanisms.
These hydrodynamic effects can be important for determining
the final light
element abundance yields in our IBBN models.

We find considerable
hydrodynamic-expansion-induced destruction of $^7$Be ($^7$Li) only
for IBBN models which employ low average $\Omega_b$ ($\Omega_b$
$^<_{\sim}$ $0.013h^{-2}$). At higher $\Omega_b$ there is little
hydrodynamic influence on the $^7$Li yield. Our results show that it
is not possible to invoke spherically condensed fluctuations to
circumvent the upper bound on $\Omega_b$ derived from HBBN.

Even for low $\Omega_b$, the only way to reconcile IBBN abundance yields with
observational limits is to fine tune fluctuation parameters. This
leads to the conclusion that any process in the early universe which
generates non-linear fluctuations prior to the epoch where $T\approx 100$
keV may be subject to nucleosynthesis constraints.

\vskip 0.15in
\centerline{\bf Acknowledgements}
\vskip 0.08in

We wish to thank C. R. Alcock, W. A. Fowler, B. S. Meyer, L. H. Kawano, and M.
S.
Smith for useful conversations and helpful suggestions. This
work was supported in part by NSF Grant PHY91-21623 and IGPP LLNL
Grant 93-22. It was also performed in part under the auspices of the
U.S. Department of Energy by the Lawrence Livermore National
Laboratory under contract number W-7405-ENG-48 and DoE Nuclear
Theory Grant SF-ENG-48.

\vfil\eject

\centerline{\bf References}

\noindent
Alcock, C. R., Fuller, G. M., \& Mathews, G. J. 1987, ApJ, 320, 439

\noindent
Alcock, C. R., Dearborn, D. S. P., Fuller, G. M., Mathews, G. J.,
\& Meyer, B. 1990,

Phys. Rev. Lett., 64, 2607

\noindent
Applegate, J. H., Hogan, C. J., \& Scherrer, R. J. 1987, Phys. Rev., D35, 1151

\noindent
Applegate, J. H. 1988, Phys. Rep., 163, 141

\noindent
Applegate, J. H., Cowan, J. J., Rauscher, T., Thielemann, F. K.,
\& Wiescher, M. 1993,

ApJ, submitted

\noindent
Boyd, R. N., Mitchell, C. A., \& Meyer, B. S. 1993, Phys. Rev. C, in
press

\noindent
Caughlan, G. R., \& Fowler, W. A. 1988, Atomic Data Nucl. Data, 40,
291

\noindent
Dearborn, D. S. P., Schramm, D. N., \& Steigman, G. 1986,
ApJ, 302, 35

\noindent
Dolgov, A., \& Silk, J. 1992, BERKELEY-CfPA-TH-92-04, preprint

\noindent
Epstein, R. I., \& Petrosian, V. 1975, ApJ, 197, 281

\noindent
Fowler, W. A. 1993, Phys. Rep., in press

\noindent
Fuller, G. M., Mathews, G. J., \& Alcock, C. R. 1988, Phys. Rev., D37, 1380

\noindent
Gnedin, N. Y., \& Ostriker, J. P. 1992, ApJ, 400, 1

\noindent
Heckler, A. , \& Hogan, C. J. 1993, Phys. Rev. D, submitted

\noindent
Jedamzik, K., \& Fuller, G. M. 1993, ApJ, in press

\noindent
Jedamzik, K., Fuller, G. M., Mathews, G. J., \& Kajino, T. 1993,
ApJ, in press

\noindent
Kawano, L. H. 1992, FERMILAB-PUB-92/04-A, preprint

\noindent
Kolb, E. W., \& Turner, M. S. 1989, in {\sl The Early Universe}, Addison-Wesley

\noindent
Kurki-Suonio, H., \& Matzner, R. 1989, Phys. Rev., D39, 1046

\noindent
Kurki-Suonio, H, \& Matzner, R. 1990, Phys. Rev., D42, 1047

\noindent
Kurki-Suonio, H., Matzner, R. A., Centrella, J., Rothman, T., \& Wilson, J. R.
1988,

Phys. Rev., D38, 1091

\noindent
Malaney, R. A., \& Fowler, W. A. 1988, ApJ, 333, 14

\noindent
Malaney, R. A., \& Mathews, G. J. 1993, Phys. Rep., in press

\noindent
Mathews, G. J., Meyer B. S., Alcock C. R., \& Fuller, G. M. 1990,
ApJ, 358, 36

\noindent
Meyer, B. S., Alcock, C. R., Mathews, G. J., \& Fuller, G. M. 1991,
Phys. Rev., D43, 1079

\noindent
Schramm, D. N., \& Wagoner, R. V. 1977, Ann. Rev. Nucl. Part. Sci., 27, 37

\noindent
Smith, M. S., Kawano, L. H., \& Malaney, R. A. 1993, ApJS, 85, 219

\noindent
Terasawa, N., \& Sato, K. 1989a, Prog. Theor. Phys., 81, 254

\noindent
Terasawa, N., \& Sato, K. 1989b, Phys. Rev., D39, 2893

\noindent
Terasawa, N., \& Sato, K. 1989c, Prog. Theor. Phys., 81, 1085

\noindent
Terasawa, N., \& Sato, K. 1990, ApJ Lett., 362, L47

\noindent
Yang, J., Turner, M. S., Steigman, G., Schramm, D. N., \& Olive, K. A. 1984,
ApJ,

281, 493

\noindent
Wagoner, R. V., Fowler, W. A., \& Hoyle, F. 1967, ApJ, 148, 3

\noindent
Wagoner, R. V. 1973, ApJ, 197, 343

\noindent
Walker, T. P., Steigman, G., Schramm, D. N., Olive, K. A., \& Kang, H. 1991,
ApJ,

376, 51

\vfil\eject

\centerline{\bf Figure Captions}

\noindent
{\bf Figure 1:} Schematic representation of fluctuation amplitude
$\Delta (x)$ plotted against position $x$. Initial fluctuation
separation $2l_{100}^s$ and width $2a_{100}$ are shown.

\noindent
{\bf Figure 2:} A plot of the computed ${}^7{\rm Li/H}$ ratio
(dashed line) and $^4$He mass fraction $Y_p$ (solid line) against
the number of baryon zones employed in the calculation. Results on
this plot are for a fluctuation with $\Lambda =1.25\times 10^6$,
$a_{100}=0.5$ m, and $l_{100}^s=10$ m. For this set of calculations
we take $\Omega_b=0.025h^{-2}$.

\noindent
{\bf Figure 3:} The ${\rm log}_{10}$ of the fluctuation amplitude
${\rm log}_{10}\Delta_i$ is shown as a function of cell radius
$r_{100}$ for protons (solid line), neutrons (dotted line),
$^2$H (long-dashed line), $^3$He (short-dashed-dotted line),
$^4$He (short-dashed line), $^7$Li (long-dashed-dotted line),
and $^7$Be (short-dashed line). Note that the fluctuation amplitude
for $^4$He is orders of magnitude larger than that for $^7$Be.
Shown are ${\rm log}_{10}\Delta_i$ profiles at epochs $(T/{\rm
MeV})=$ 100, 5, 2.5, 1.0, 0.25, 0.10, 0.05, 0.025, 0.01.
An overall average $\Omega_b=0.025h^{-2}$ was used in this
calculation. The initial fluctuation characteristics assumed for
this calculation are $\Lambda =3\times 10^7$, $l_{100}^s=0.5$ m,
and $l_{100}^s/a_{100}=25$.

\noindent
{\bf Figure 4:} Similar to Figure 3 but with $\Omega_b=0.25h^{-2}$,
and initial fluctuation parameters $\Lambda =3\times 10^7$,
$l_{100}^s=50$ m, and $l_{100}^s/a_{100}=25$. In this plot
${\rm log}_{10}\Delta_i$ profiles are given at epochs $(T/{\rm
MeV})=$ 100, 1.0, 0.25, 0.10, 0.075, 0.050, 0.025, 0.01, 0.008.

\noindent
{\bf Figure 5:} Same as Figure 4 but with parameters
$\Omega_b=0.0125h^{-2}$, $\Lambda =1.25\times 10^6$, $l_{100}^s=14$
m, and $l_{100}^s/a_{100}=20$.

\noindent
{\bf Figure 6:} Similar to Figure 4 but with parameters
$\Omega_b=0.25h^{-2}$, $\Lambda =2\times 10^9$, $l_{100}^s=50$ m,
and $l_{100}^s/a_{100}=100$. In this figure ${\rm log}_{10}\Delta_i$
profiles are given for the fluctuation core region ($r_{100}\leq 10$
m) only.

\noindent
{\bf Figure 7:} Same as Figure 6 but with parameters
$\Omega_b=0.0083h^{-2}$, $\Lambda =1.25\times 10^7$, $l_{100}^s=50$
m, and $l_{100}^s/a_{100}=1000$. The ${\rm log}_{10}\Delta_i$
profiles
are shown for fluctuation core region ($r_{100}\leq 3$ m) only.

\noindent
{\bf Figure 8:} Average nucleosynthesis yields are shown as a
function of initial fluctuation cell radius $l_{100}^s$ in meters.
The upper left hand panel gives the $^4$He mass fraction ${\rm Y}_p$,
while the lower left hand and upper right hand panels give number
fractions relative to hydrogen for $^7$Be, $^7$Li, $^3$He, $^2$H,
and $^3$H as indicated. The two lower right hand plots give the
average $^7$Be/H ratios (at $T=20$ keV and $T=5$ keV) and the
average neutron-to-photon ratio (at $T=20$ keV) as a function of
$l_{100}^s$. These calculations were performed for fixed $\Lambda
=3\times 10^7$ and $l_{100}^s/a_{100}=25$, and assume an average
$\Omega_b=0.25h^{-2}$.

\noindent
{\bf Figure 9:} Average nucleosynthesis yields are shown as a
function of initial fluctuation Gaussian width $a_{100}$ in meters.
The panels are as in Figure 8. These calculations assume average
$\Omega_b=0.25h^{-2}$ and initial fluctuation cell radius
$l_{100}^s=50$ m. The initial fluctuation density contrast $\Lambda$
satisfies $\Lambda\geq 10^6$.

\noindent
{\bf Figure 10:} Same as Figure 8 but with average
$\Omega_b=0.025h^{-2}$ and initial fluctuation parameters $\Lambda
=1.25\times 10^6$, and $l_{100}^s/a_{100}=20$.

\noindent
{\bf Figure 11:} Same as Figure 9 but with average
$\Omega_b=0.025h^{-2}$ and initial fluctuation parameters $\Lambda
=1.25\times 10^7$ and $l_{100}^s=10$ m.

\noindent
{\bf Figure 12:} Same as Figure 8 but with average
$\Omega_b=0.0125h^{-2}$ and initial fluctuation parameters $\Lambda
=1.25\times 10^6$ and $l_{100}^s/a_{100}=20$.

\noindent
{\bf Figure 13:} Same as Figure 9 but with average
$\Omega_b=0.0125h^{-2}$ and initial fluctuation parameters
$\Lambda =1.25\times 10^7$ and $l_{100}^s=24$ m.

\noindent
{\bf Figure 14:} Same as Figure 8 but with average
$\Omega_b=0.005h^{-2}$ and initial fluctuation parameters $\Lambda
=1.25\times 10^6$ and $l_{100}^s/a_{100}=20$.

\noindent
{\bf Figure 15:} Average $^7$Be/H abundance (in units of $10^{-8}$)
is shown as a function of cosmic temperature $T$ in keV for five
calculations. Correspondence between individual curves and
calculation techniques is as indicated in the key. All results employ
a fluctuation with parameters $\Lambda =2\times 10^7$,
$l_{100}^s=50$ m, $l_{100}^s/a_{100}=20$, and average
$\Omega_b=0.25h^{-2}$.

\noindent
{\bf Figure 16:} $^7$Li/H and $^7$Be/H in units of $10^{-8}$ are
plotted against temperature $T$ in keV. Solid lines give results for
a calculation with full hydrodynamic effects. Dashed lines are for a
calculation without hydrodynamic expansion effects. The calculations
employ $\Omega_b=0.025h^{-2}$ and have initial fluctuation parameters
$\Lambda =1.25\times 10^6$, $l_{100}^s=12$ m, and
$l_{100}^s/a_{100}=20$.

\noindent
{\bf Figure 17:} Same as Figure 16 but for calculations with
$\Omega_b=0.0125h^{-2}$. Fluctuation parameters are the same as in
Figure 16 except that $l_{100}^s=14$ m.

\end